\documentclass[preprint2]{aastex}
\usepackage[T1]{fontenc}
\usepackage{graphics,graphicx,epsfig}
\usepackage{amssymb,amsmath,amsfonts,amssymb,subfigure}
\usepackage{mathrsfs}

\shorttitle{Evolution of a dwarf satellite galaxy embedded in a SFDM halo}
\shortauthors{V.H. Robles, V. Lora et al.}

\begin{document}


\title{Evolution of a dwarf satellite galaxy embedded in a scalar field dark matter halo}


\author{Victor H. Robles$^{1}$\altaffilmark{*}, V. Lora$^{2}$\altaffilmark{**}, T. Matos$^{1}$, F.J. S\'anchez-Salcedo$^{3}$}
\affil{$^1$Departamento de F\'isica,Centro de Investigaci\'on y de Estudios Avanzados del IPN, 07000 D.F., M\'exico} 
\affil{$^2$Astronomisches Rechen-Institut, Zentrum f\"{u}r Astronomie der Universit\"{a}t Heidelberg, M\"{o}nchhofstr. 12-14, 69120 Heidelberg, Germany} 
\affil{$^3$Instituto de Astronom\'{\i}a,Universidad Nacional Aut\'onoma de M\'{e}xico, AP 70-264, 04510 D.F., M\'{e}xico}



\altaffiltext{*}{E-mail:\tt{vrobles@fis.cinvestav.mx}}
\altaffiltext{**}{E-mail: \tt{vlora@ari.uni-heidelberg.de}}


\begin{abstract}
The cold dark matter (CDM) model has two unsolved issues: simulations overpredict the satellite abundance around the Milky Way (MW) and it disagrees with observations of the central densities of dwarf galaxies which prefer constant density (core) profiles.One alternative explanation known as the scalar field dark matter (SFDM) model, assumes that the dark matter is a scalar field of mass($\sim 10^{-22}$ eV/$c^2$); this model can reduce the overabundance issue due to the lack of halo formation below a mass scale of $\sim 10^8$M$_{\odot}$ and successfully fits the density distribution in dwarfs. One of the attractive features of the model is predicting core profiles in halos, although the determination of the core sizes is set by 
fitting the observational data. We perform \textit{N}-body simulations to explore the 
influence of tidal forces over a stellar distribution embedded in a SFDM halo orbiting a MW-like SFDM host halo with a disk. Our simulations intend to test the viability of SFDM as an alternative  model by comparing the tidal effects that result in this paradigm with those obtained in CDM for similar mass halos. We found that galaxies in subhalos with core profiles and high central densities survive for 10 Gyr. The same occurs for galaxies in low density subhalos located far from the host disk influence, whereas satellites in low density 
DM halos and in tight orbits can eventually be stripped of stars.
We conclude that SFDM shows consistency with results from CDM for dwarf galaxies, but naturally offer a possibility to solve the missing satellite problem.

\end{abstract}





\section{Introduction}

The Lambda Cold Dark Matter ($\Lambda$CDM) paradigm  has been very successful
in explaining the structure formation on large scales. One of its predictions is a universal
density profile for the dark matter halos. Navarro, Frenk \& White (1997, NFW) suggested
a simple formula to describe these density profiles, which presents a divergent inner
profile ($\rho(r)\propto r^{-1}$) \citep{die05}. Another prediction of the $\Lambda$CDM model is the number of 
subhalos per unit mass around the host galaxy. Both predictions have been challenged
on scales of dwarf galaxies. In fact, a significant fraction of the rotation curves of low surface brightness (LSB)
galaxies and dwarf irregular galaxies are better fitted using dark halos with a density core 
($\rho (r) \propto r^{0}$) \cite[]{wil04,gen04,rea06,str06,deb08,oh08,tra08,wal11,agn12,pen12,
sal12,lor13}. 
However, the case for cores in Milky Way (MW) satellites is still debated. For instance, 
\cite{str14} mentioned that the data of the Sculptor dwarf spheroidal are consistent  
when an NFW dark matter halo is assumed. Moreover, for dwarf galaxies in the field or Andromeda the information is 
about the dark matter mass and not the density profiles, so a direct determination of the slope in the density profile is not possible.
Regarding the prediction in the number of subhalos, it turns out that the standard CDM model overpredicts the number
of dwarf satellite galaxies in the MW and M31.
This disagreement is usually referred to as the ``missing satellite problem'' 
\cite[]{kly99,moo99,goe07,sim07,bel10,mac12,gar14b}. Although the detection of ultra faint galaxies within the MW halo
has reduced the missing satellite problem (e.g. Simon \& Geha 2007), a recent study by \citet{iba14}  of the distribution of satellites around the MW and M31 suggests they have specific  alignments forming planes that are not found in current CDM simulations. 
Independently of this potential issue, the central densities of MW dSph galaxies are required to be significantly lower than the densities of the largest subhalos found in collisionless DM simulations to agree with current data\citep{boy11,gar14}.
Indeed, CDM simulations of the Aquarius Project \citep{nav10} suggest that the MW size halos should
inhabit at least eight subhalos with maximum circular velocities exceeding $30$ km s$^{-1}$,
while observations indicate that only three satellite galaxies of the MW possess halos with maximum circular velocities $>30$ km s$^{-1}$. This discrepancy is known as the ''too big to fail''  problem.

It has been argued that the physics of baryons must be included in order to 
make a fair interpretation of observations on scales of MW subhalos.
For instance, mass outflows given by supernova explosions could transform
a cusp into a core in some field dSph galaxies at the present time.
The missing satellite discrepancy may be explained as a consequence of 
gas reionization that quenched the star formation in halos with
maximum circular velocity less than $20$ km s$^{-1}$, leaving hundreds of small mass halos without stars \citep{boy14}. In principle using gravitational lensing techniques could confirm the existence of these halos.  
However, there is still no consensus on whether mass outflows and
reionization can explain the observed properties of the MW satellite galaxies \citep{pen12,oka08}.
Additionally, it seems there could be a numerical code dependence when interpreting the 
results obtained from simulations \cite[]{sca01,sca13}.

More recently, it was noted that CDM predicts massive subhalos with central densities higher than those found in satellite galaxies, meaning that there are DM subhalos that are massive but host no satellite galaxies \cite[]{boy11,ras12,tol12}. All these problems might be related and share a common solution. The way they are correlated usually depends on the dark model paradigm 
\cite[]{don13,roc13}, or gravity model \cite[]{mil10,mac13}, but a general fact is that solving one of these issues provides clues to the solution for the other issues. 
It it worth mentioning that although supernovae explosions seem to play a crucial role in forming cores in field dwarf galaxies and more massive systems where the gas is recycled to continue the star formation \cite[]{mas08,bro11,pon12,gov10,gov12,gar13},  it is unclear that the same feedback implementation works in satellite galaxies where the gas content is negligible and their stellar populations are mostly dominated by old stars. In this sense, dark matter models where the core formation is through DM properties and not by the specifics of astrophysical processes are still viable alternative solutions. 

One of these alternative models is the scalar field dark matter (SFDM) model. 
The idea was first considered by \citet{sin94} and independently introduced by \citet{guz00}.
In the SFDM model the main hypothesis is that the dark matter is a self-interacting real scalar field of a small mass ($m\sim 10^{-22}$ eV/c$^2$) that condensates forming Bose-Einstein condensate (BEC) ``drops''  \cite[]{mag12a, lor12}. We interpret these BEC drops as the halos
of galaxies \citep{mat01} such that the DM wave properties and the Heisenberg uncertainty principle stop the DM phase-space density from growing indefinitely. These properties automatically avoid the divergent density (cuspy) profiles in DM halos and reduce the number of small satellites due to the mass cut-off in the power spectrum \cite[]{hu00,mar14}.
For this typical mass, it follows that the critical temperature of condensation of the scalar field is T$_\mathrm{crit}\sim m^{-5/3}\sim$TeV,  thus, BEC drops can be formed very early in the universe. There have also been numerous studies that analyzed the behavior of the scalar field at large scales \cite[]{sua11,sua13,mat01,mat07,mag12b,hu00,har11,cha11,ber92}, concluding that it reproduces the successes of the CDM model at those scales. 

One straightforward and universal prediction of the wave properties of this model is that DM halos have core profiles since their initial formation \cite[]{rob13,sua13}. If halo distributions are flat from the beginning, then 
strong feedback blowouts are not required 
to produce low density distributions in DM-dominated systems. Some other consequences of this particular feature have been explored in different contexts; to fit rotation curves in LSB and dwarf galaxies \cite[]{rob13,har11,cha11,lor12,lor14}, and to make strong lensing analyses \cite[]{rob13b, gon13}. 

All these successes of the model have motivated us 
to test the model further in order to know if 
it can be regarded as a serious DM candidate in the universe; conducting these tests is necessary for models 
whose DM properties are quite different from those of the standard classical particle description. Thus, here we study
the evolution of the stellar component of satellite dwarf galaxies embedded in SFDM halos orbiting within an SFDM MW-size host halo.
Our study provides constraints on both the final stellar distribution of dSphs and the survival of faint or ultra faint systems. 

We pursue this task by studying the conditions under which tidal disruption may occur in the SFDM model. Previous studies have shown, using empirical core-like density profiles for DM halos, that tidal disruption can be more important than in halos with NFW profiles,  especially if they pass
close to the galactic disk (see Klimentowski et al. (2009) and Pe\~narrubia et al. (2010) for collisionless simulations).  However, until now there have not been studies addressing whether the tidal effects are strong enough to completely remove the stars in classical and ultra faint dwarf galaxies hosted by  BEC halos. The present  work aims to investigate this issue through a series of simulations of a stellar component described by a Plummer profile when it is embedded in a SFDM subhalo subject to the influence of a SFDM host halo with a disk component. We also conducted simulations without the disk to compare its effect on the stars.

The article is organized as follows: in Section 2 we explain the SFDM model and present the density profiles to be used in the simulations. Section 3 describes the simulations, Section 4 contains our results and discussions of the satellite galaxy evolution, and Section 5 presents our conclusions.

\section[]{The baryonic components}

\subsection{The dSph stellar component}
\label{sec:dwarf}

The dSph galaxies have low luminosities ($L_V\sim10^2-10^7$L$_{\odot}$) and very large dynamical 
mass-to-light ratios $M/L\gtrsim10$, which translate into a large amount of DM \citep{mun05,str06,gil07,
mat08,wal09,mac10,wol10}. Nevertheless, we detect the galaxies because of the stars and, in fact, using them as tracers of the potential gives  us information about their potential well.
Here we use a Plummer density profile \citep{plu11} for the stellar component of the dSph, where the mass density profile is given by
\begin{equation}
 \rho(r)= \frac{3M_{*}}{4 \pi r_{p}^3} \left( 1+\frac{r}{r_{p}} \right) ^{-5/2} \hbox{ ,}
\end{equation}
where $M_{*}$ is the mass of the stellar component, and $r_p$ is the Plummer radius. 
One should note that $r_p$ can be related to the half-mass-radius $r_h$ through  $r_{h}=1.3r_{p}$. 
In our simulations, we have set a half-mass-radius of $200$ pc, and a stellar mass of $M_{*}\simeq7.3\times10^5$~M$_{\odot}$, 
motivated by the typical values for Draco, which is one of 
the classical dSph galaxies  and also one of the least luminous satellites (e.g., see \citeauthor{mar08} \citeyear{mar08} and 
\citeauthor{oden01} \citeyear{oden01}, for Draco).

\subsection{The MW Disk Component}
\label{sec:mw_disc}
In some of our simulations we include the potential of the MW's
baryonic disk, which we model with a Miyamoto-Nagai potential 
\citep{miy75}

\begin{equation}
\Phi_d(R,z) = -\frac{G M_d}{\sqrt{ R^2 + (a + \sqrt{z^2+b^2})^2}} \hbox{.}
\end{equation}
In the latter equation, $M_d$ stands for the mass of the disk, and $a$ and $b$ stand for the 
horizontal and vertical scalelengths, respectively. We have set the mass of the disk to 
$M_d=7.7\times10^{10}$~M$_{\odot}$, and the scalelengths to $a=6$~kpc, and $b=0.3$~kpc. 

\subsection[]{The dark matter component}
\label{sec:dm_dwarf}

\subsubsection{The dSph DM component}

\begin{figure*}
  \centering
  \includegraphics[width=1.0\textwidth]{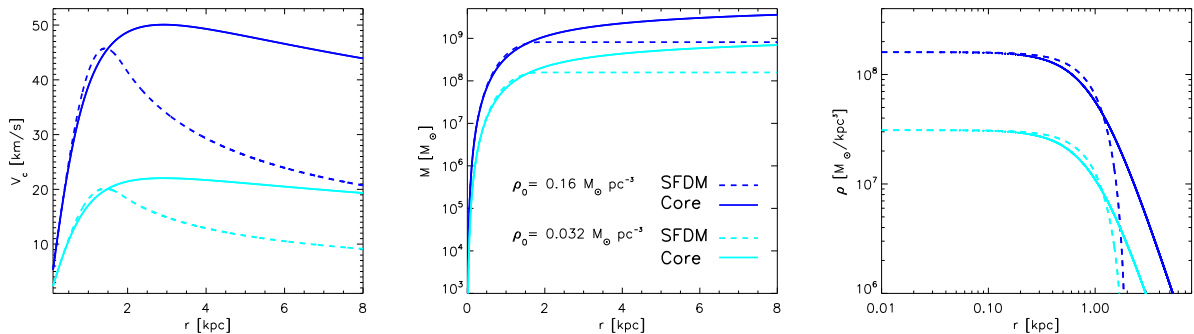}
  \caption{The circular velocity(left), mass(center), and density profile(right) associated to the two SFDM halos of the dwarf galaxy (dashed 
black line (blue in the online version) is model A: $0.16$~M$_{\odot}$~pc$^{-3}$, and dashed gray line (cyan in the online version) is model 
  B: $0.031$~M$_{\odot}$~pc$^{-3}$). The corresponding core DM models are shown with solid lines with their respective colors for comparison.}
  \label{fig:halos}
\end{figure*}
If DM is composed of scalar particles with masses $m \ll$ 1 eV/c$^2$, the galactic halos have very large occupation numbers and the field behaves as a classical field that obeys the Klein-Gordon equation. For SFDM halos, the Newtonian limit is enough to describe them. From the fits to the rotation curves of DM-dominated systems, it has been found that the SFDM halos of DM-dominated galaxies are well described with the ground state \citep{guz06,boh07,rob12,mart14a}. However, the larger the galaxy the more important are the effects of the non-condensed states on the mass profile. The latter means that galaxies that have RC that remains flat even at large radii are better described by adding the excited state contributions \cite[]{ber10,rob13}. This suggests that excited states are relevant to describe MW size systems.  Their relevance in dwarfs is out of the scope of this work but see, for instance, \cite{mart15}. 
In this work we will then consider the base state to describe the dwarf DM halos.

Following the hypotheses mentioned above for the SF and using the temperature corrections to one loop for the scalar field, Robles \& Matos (2013) found that after the phase transition that happens in the early universe, the field rolls down to a new minimum of the potential and reaches those values where it will remain. The structures will grow and eventually form the SF halos. Assuming the field is at the minimum, the authors derive an analytical solution for a static spherical configuration that allows the presence of excited states\footnote{We refer the reader to the mentioned work for details on the calculation of the density profile of a SFDM halo given that the mathematical details are already described in that work.}. 
What they found is that for a SFDM halo in the state $j$ its density profile is given by 
\begin{equation}
\label{eq:sfdm}
 \rho_j(r)=\rho_{0,j}\frac{\sin^2(k_j r)}{(k_j r)^2} \hbox{ .}
\end{equation}

In the latter equation, $\rho_{0,j}$ is the central mass density, $k_j \equiv j \pi / R_h$, $j$ is a positive integer that 
identifies the minimum excited state needed to fit the data of a galaxy, 
$R_h$ is a scalelength that is determined from observations and its a free parameter; fitting data for a given galaxy provides 
values for both parameters, the scale $R_h$ and the central density, and the same occurs using eq.(\ref{eq:core}) but for its own parameters.
There is not observational evidence that determines how far the halo should extend, but 
we do expect that halos spread at least enough to cover up to the outermost measured data. 
It follows that if galaxies have stellar distributions mostly concentrated in the center with possibly some gas surrounding them,
then $R_h$ would be larger than the radius where most of the stars are confined. 
Based on the trend from the fits of works that use scalar field dark matter halos\cite[]{rob12,rob13,lor12,har11} where 
the scale radius complies with the above condition, we may take the $R_h$ to be a truncation radius such 
that $\rho(r)=0$ for $r$>$R_h$ and for all $r\leq$ $R_h$ the density is given by eq.(\ref{eq:sfdm}).

We recall that scalar field configurations in excited states are characterized by nodes, thus, for a bounded 
configuration the ground state has no nodes and corresponds to $j=1$, the first excited state has one node and it is associated with $j=2$, 
the second has two nodes, and so on. We remark that from this interpretation, if we are dealing with a field configuration in the ground state 
corresponding to zero nodes, then eq.(\ref{eq:sfdm}) has no oscillatory behavior; only those configurations in excited states 
have oscillations.

From equation (\ref{eq:sfdm}) we obtain the mass and rotation curve velocity profiles given by
\begin{eqnarray}
M(r) &=& \frac{4 \pi  \rho_{0,j}}{k_j^2} \frac{r}{2} \biggl(1-\frac{\sin(2 k_j r)}{2 k_j r} \biggr), \\
V^2(r) &=& \frac{4 \pi G \rho_{0,j}}{2 k_j^2} \biggl(1-\frac{\sin(2 k_j r)}{2 k_j r} \biggr) \label{vel}. 
\end{eqnarray}
respectively.
 \citet{diez14} reported that MW dSphs that are within SFDM halos in the ground state are well described with truncation radii in the range $\sim$ 0.5$-$2 kpc, they mentioned that  a common value larger than 5 kpc is disfavored by the dynamics of dSphs provided they are in SFDM halos where only the ground state is taken into account.  \cite{mart15} extended this result to account for higher energy states of the scalar field in the associated SFDM halos of the MW dSphs  and found that the 
the stellar distribution lies inside the region where the ground state of the SF is mostly confined ($ \sim$ 0.5 -1.5 kpc), also, the presence of the first excited state does not substantially affect the innermost dark matter configuration but it does allow for the possibility of a larger truncation radius ($\sim$ 5 kpc), nevertheless current data in the dwarfs analyzed are insufficient to conclude the existence of other states in SFDM halos of MW dSphs and hence determining precisely the halo radius. For our generic analysis of a typical dwarf we will then consider a value of $R_h$ =2 kpc consistent with the results suggested by the above independent analyses, additionally, notice that most of the stellar component resides inside 1 kpc in most dSphs, where the dark matter distribution is not substantially  modified by the precise halo radius that is considered as shown in \cite{mart15}.

For the parameters of the dwarf DM halo we then adopt the values $j=1$ and a typical radius of $R_h=2$ kpc. Notice that for the base state $j=1$ there is no oscillatory behavior in the RC (Figure 1) contrary to what the case with excited states (Figure 2). 

For the dwarf central density ($\rho_{0,1}$), we select two different values that encompass the range of masses found in dwarfs, $0.16$~M$_{\odot}$~pc$^{-3}$
(model A) and a less  massive one with  $0.031$~M$_{\odot}$~pc$^{-3}$ (model B).  

In Figure \ref{fig:halos}, the dashed lines show the circular velocity, 
mass, and density associated with the SFDM halos of models A (black (blue online)) and B (gray (cyan online)). 
The corresponding SFDM dwarf core radius (defined as the radius at which the central density drops a factor of two) is $\sim750$~pc for both A and B models and its presence is distinctive prediction of the model. 

To compare the SFDM profiles with other cored profiles, 
we also consider the following profile \citep{pen10} 
\begin{equation}
\label{eq:core}
 \rho(r)=\frac{\rho_0}{(1+(r/R_{s})^2)^{3/2}} \hbox{ .}
\end{equation}
For both A and B models, we set the scale radius $R_s=1$~kpc (see solid lines of Figure~\ref{fig:halos}).

For our mass models, the mass of the dark halo enclosed at $R_{h}=2$ kpc
lies in the range $10^{8}$-$10^{9}$ M$_{\odot}$. The resulting mass-to-light ratios represent
DM dominated dSphs. For instance,
the mass-to-light ratios of dSphs ([M/L]$_{half}$) in the MW range from 
$\sim 7$~M$_{\odot}$/L$_{\odot}$ (Leo I, Fornax) to $\sim(10^3)$~M$_{\odot}$/L$_{\odot}$ 
(UMa II, Seg, UMaI) \citep{col14}. In particular, Draco has a very low luminosity but a high estimated total mass within the tidal radius of $M(r_t)=2.2-3.5\times10^7$~M$_{\odot}$ \citep{oden01}, this leads to a 
high mass-to-light ratio of $(M/L)_{i}\simeq92-146$\citep{irw95,ama95}.

\subsubsection{The MW DM component}
\label{sec:dm_mw}

\begin{figure*}
  \centering
  \includegraphics[width=1.0\textwidth]{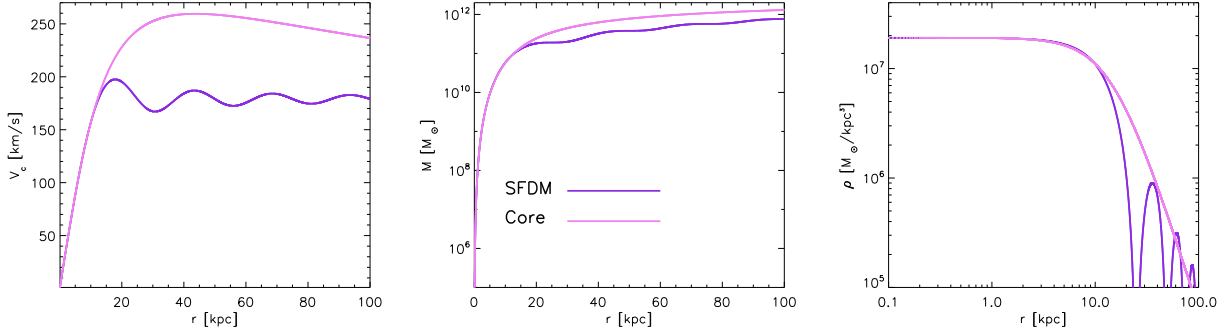}
  \caption{From left to right are the cicular velocity, mass, and density profiles used for: the MW's SFDM halo model (black lines (purple in the online version)), 
and the cored DM halo (gray lines (pink in the online version).}
  \label{fig:haloMW}
\end{figure*}
In the SFDM model, the fluctuations are expected to grow faster than in the standard model 
\citep{sua11}, implying that galaxies are fully formed at large redshifts. In fact, some recent 
high redshift observations suggest the existence of well formed galaxies very early in the universe
\citep{col09,fin13,cha14}. From the results of hydrodynamical CDM simulations that model the MW, one sees that its dark and luminous matter do not substantially change since $z \lesssim 2$  
($\sim 10$~Gyr ago) \citep{die07,gov07,kli09,pen10,kas12}. Then, for the initial conditions, we can assume 
a host with similar parameters that reproduce current MW data. We found that using $\rho_{0,4}=0.0191$~M$_{\odot}$~pc$^{-3}$, $j=4$, and $R_{h}=100$~kpc (see thick solid lines (purple in the online version) in Figure~\ref{fig:haloMW}) in Equation~(\ref{eq:sfdm}) gives a good representation to the MW DM in the SFDM model. 
Although a detailed analysis of the MW with the SFDM is out of the scope of this work, we obtained the quoted values following the usual procedure of estimating the parameters that model our neighborhood, that is,  we search the parameters consistent with the circular velocity in the solar neighborhood and the Oort constants, we find a velocity $\sim$ 200 km/s at 8.5 kpc and constants A=15.5 km s$^{-1}$ kpc$^{-1}$ and B=-14.4km s$^{-1}$ kpc$^{-1}$ similar to previous works\citep{feast97}, we did the estimation when the disk is present and obtained the above values for the SFDM halo and the disk parameters reported following eq. (2). 

For the MW's cored DM profile (Equation~\ref{eq:core}), we set $R_{s}=15$~kpc. The corresponding circular velocity, 
mass and density of the cored DM halo, are shown with gray lines (pink in the online version) in Figure~\ref{fig:haloMW}.
It has to be noted that, for both (SFDM and cored) MW halos, the core radius is $\sim11.5$~kpc, and that the mass estimations 
within 100 kpc are comparable. Therefore, the DM profiles are not identical
but the total mass enclosed at the halo radius is the same. Given that all our satellites have orbits inside this radius,  we choose $R_h$=100 kpc, principally because our main focus is to study the tidal stripping of the stellar component of these satellites, whose apocenters never become larger than 100 kpc during the simulation, as the MW dark matter mass outside this radius is not essential to our study we can truncate the halo at that point,  it then follows a Keplerian decay for larger radii. The wiggles found in the halo and shown in 
Figure \ref{fig:haloMW} are also a particular difference of this SFDM profile with respect to other core models.  

\section[]{Simulations}
\label{sec:code}

We simulate the evolution of the stellar component of
the dwarf galaxy, which is embedded in a rigid SFDM 
halo potential using the $N$-body code \scriptsize {SUPERBOX} \normalsize \citep{fellhauer00}. 
\scriptsize {SUPERBOX} \normalsize is a highly efficient particle-mesh, collisionless-dynamics 
code with high resolution sub-grids. 
In our case, \scriptsize {SUPERBOX} \normalsize uses three nested grids centered in the 
density center of the dwarf galaxy. We used $128^3$ cubic cells for each of 
the grids. The inner grid is meant to resolve the inner region of the dwarf galaxy.
The spatial resolution is determined by the number of grid cells 
per dimension ($N_c$) and the grid radius ($r_{\rm grid}$). Then the side length of one 
grid cell is defined as $l=\frac{2 r_{\rm grid}}{N_c-4}$. For $N_{c}=128$, the resolution
is  $0.5$~pc.
\scriptsize {SUPERBOX} \normalsize integrates the equations of motion with a leap-frog 
algorithm, and a constant time step $dt$. We selected a time step of $dt=1$~Myr in our 
simulations. 

\begin{table*}
\centering
\caption{Parameters used in our simulations.\label{table:1}} 
\begin{tabular}{ccccccc}
\tableline\tableline
              &                      & Dwarf & Dwarf      & MW &  MW & DM\\
   Simulation & $\frac{r_{p}}{r_{a}}$& orbit & $\rho_{0}$ & $\rho_{0}$ & disk & model \\
              &                      & plane &$(10^7$ M$_{\odot}$~(kpc)$^{-3})$ & $(10^7$ M$_{\odot}$~(kpc)$^{-3})$&  &\\
\tableline 
A1 & 1/2  &x-y& 16 & 1.91 & --& SFDM \\
B1 & 1/2  &x-y& 3.1& 1.91 & --& SFDM \\
A2 & 1    &x-y& 16 & 1.91 &$\checkmark$ & SFDM\\
B2 & 1    &x-y& 3.1& 1.91 &$\checkmark$ & SFDM\\
A3 & 1/2  &x-y& 16 & 1.91 &$\checkmark$ & SFDM\\
A3$_{core}$ & 1/2  &x-y& 16 & 1.91 &$\checkmark$ & Core\\
B3,~B6\tablenotemark{a} & 1/2  &x-y& 3.1& 1.91 &$\checkmark$ & SFDM\\
B3$_{core}$,~B6$_{core}$ & 1/2  &x-y& 3.1& 1.91 &$\checkmark$& Core \\
A4,~A6\tablenotemark{b} & 1/5  &x-y& 16 & 1.91 &$\checkmark$ & SFDM\\
A4$_{core}$,~A6$_{core}$ & 1/5  &x-y& 16 & 1.91 &$\checkmark$& Core \\
B4 & 1/5  &x-y& 3.1& 1.91 &$\checkmark$ & SFDM\\
B4$_{core}$ & 1/5  &x-y& 3.1& 1.91 &$\checkmark$ & Core\\
A5 & 1/5  &45$^{\circ}$& 16 & 1.91 &$\checkmark$ & SFDM\\
B5 & 1/5  &45$^{\circ}$& 3.1& 1.91 &$\checkmark$ & SFDM\\
\tableline
\end{tabular}
\tablenotetext{a}{Simulations B6(B6$_{core}$) use the same parameters of B3(B3$_{core}$) but with a satellite stellar mass M$_{\ast} = 1 \times 10^4 \ M_{\odot}$}.
\tablenotetext{b}{Simulations A6 (A6$_{core}$) use the same parameters of A4(A4$_{core}$) but with 
a satellite stellar mass M$_{\ast}=1 \times 10^4 M_{\odot}$}.
\tablecomments{Column 1 identifies the simulation, column 2 specifies $r_{p}$/$r_{a}$ for the orbit, 
column 3 shows the plane of the orbit, next two columns give the central density for the dwarf and the MW DM halos in 
each simulation, respectively, column 6 determines if a disk is present in the Milky Way halo, and column 7 gives the DM model 
used in the simulation.}
\end{table*}
\begin{figure*}
 \centering
  \includegraphics[width=1.0\textwidth]{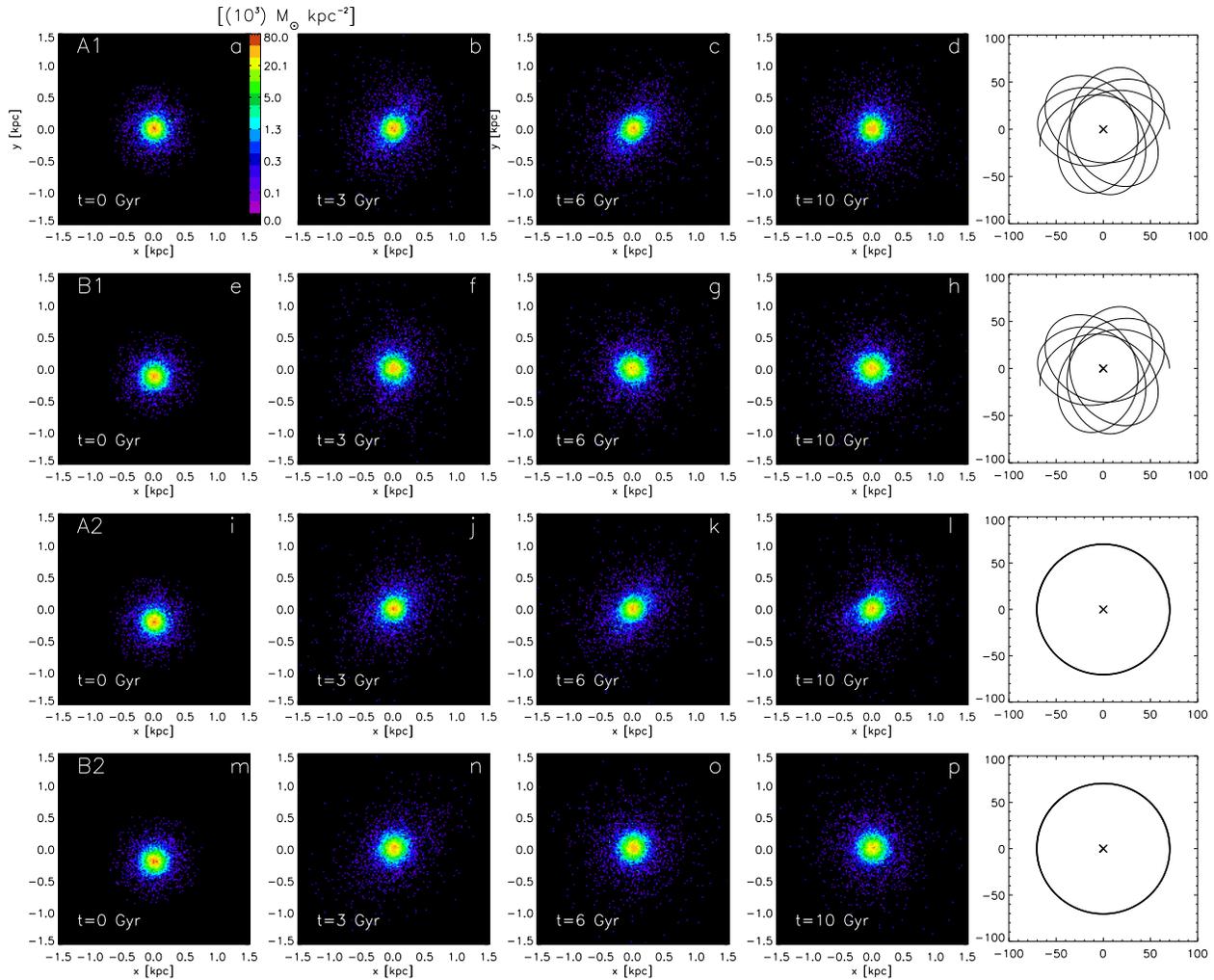}
  \caption{The surface mass density of the dwarf galaxy for models $A1,B1,A2$, and $B2$ for $t=0$, 
$3$, $6$, and $10$~Gyr, all plots are centered in the dwarf galaxy. In the last column, we show the orbit of the satellite galaxy 
around a Milky Way SFDM halo (colored version online).}
  \label{fig:vic1}
\end{figure*}

For the orbit of the dwarf galaxy, we assume an apocenter distance from the MW, $r_{a}=70$ kpc \citep{bon04} and two 
different pericenter distances  ($r_p=14$ and $35$ kpc). 
We conducted simulations with and without adding the presence of a Miyamoto-Nagai disk in the MW potential to assess the effects 
on the dwarfs due to the close encounter with the disk component.  
Our main interest is the stellar component evolution that is located deep inside the subhalo. \cite{pen10} show 
that the major effect of tidal disruption of a DM suhalo occurs in the outermost radius, while inner regions 
($\lesssim 1$ kpc) are 
less affected by tides and the density profiles are only shifted to a slightly lower value maintaining the same inner shape during 
the evolution. The evolution changes if the subhalo's pericenter is smaller than the length of the disk in the event that this component is present, meaning that when the subhalo effectively cross through the disk several times it can lose a considerable amount of its initial mass or even get destroyed if the orbit's pericenter is $\sim$1.8 kpc, however, these are rare events.
Given these results and that we consider rigid subhalos, we focus our analyses on orbits with pericenters larger than the disk scalelength avoiding direct collisions with the disk that would require a live subhalo. Since stars serve as tracers of the 
subhalo potential, any major tidal disruption of the stars would be indicative of a substantial change in the evolution of the subhalo. 
In such cases, a live halo would be needed. This happens only in one of our simulations and will not be used to   
draw the overall conclusions of this work. However, it does serve to show that our results are consistent with those in \cite{pen10}.

In our first couple of simulations, denoted by $A1$ and $B1$,  
the dwarf galaxy is embedded in the MW SFDM halo potential without including the 
baryonic MW disk component (first two rows of Figure~\ref{fig:vic1}). The dwarf 
galaxy is placed at a galactocentric distance of 70~kpc, and orbits in the 
$x-y$ plane with a $r_{p}/r_{a}=1/2$.
\begin{figure*}
  \centering
  \includegraphics[width=1.0\textwidth]{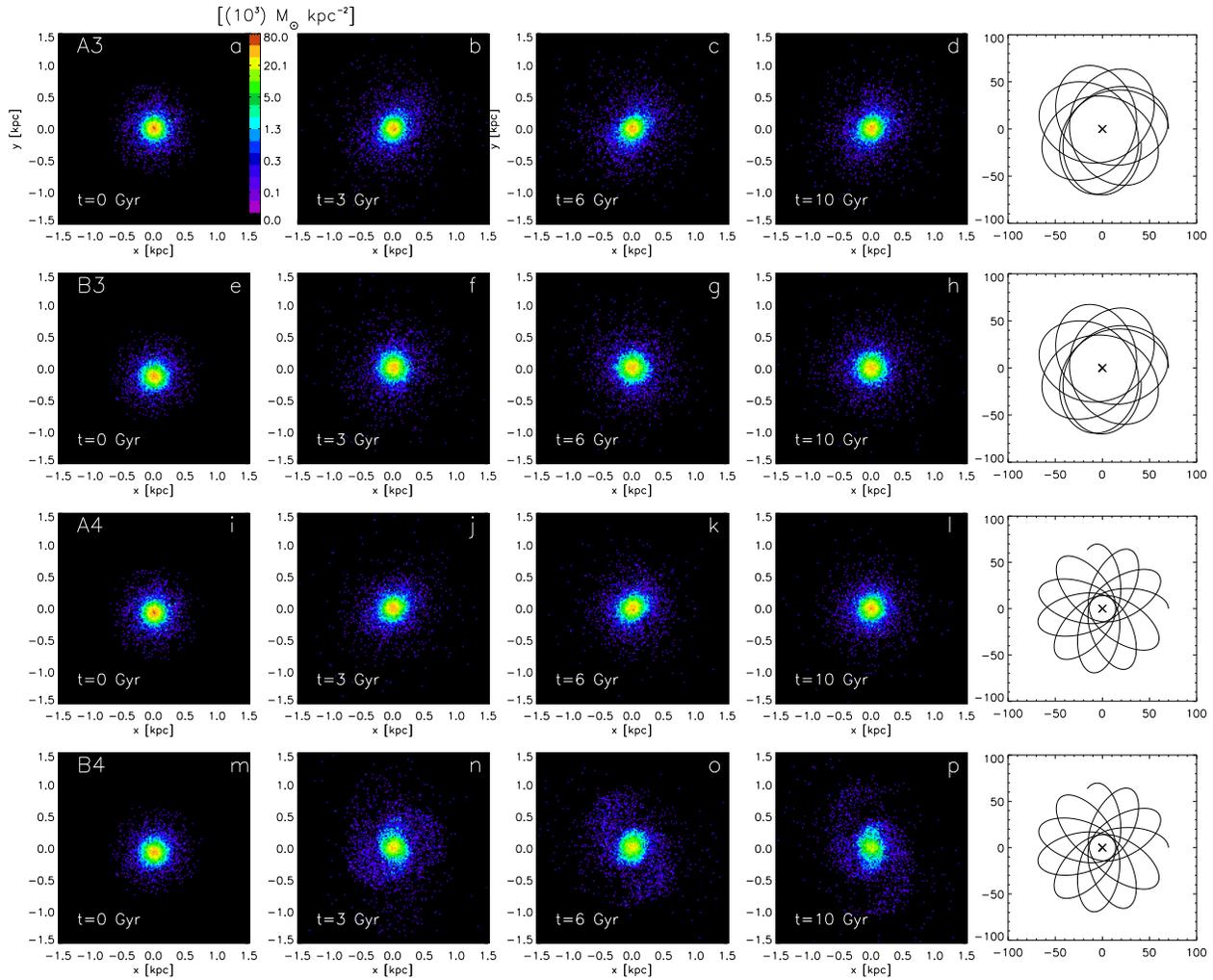}
  \caption{Same as Figure~\ref{fig:vic1}, but for models $A3-B4$(colored version online).}
  \label{fig:vic2}
\end{figure*}

In the second pair of simulations, $A2$ and $B2$, we model the dwarf galaxy 
embedded in the MW SFDM halo potential in a circular orbit ($r_{p}/r_{a}=1$), including the 
baryonic MW disk component (see last two rows of Figure~\ref{fig:vic1}).

In the third pair of simulations, $A3$ and $B3$, we model the dwarf galaxy 
embedded in the MW SFDM halo potential, with a $r_{p}/r_{a}=1/2$, but now 
we include a rigid baryonic MW disk component.  We rerun these two simulations to compare with 
the empirical profile (eq. \ref{eq:core}) referred as $A3_{core}$ and $B3_{core}$.

We observe from Figures~\ref{fig:vic1} and \ref{fig:vic2} that the dwarf galaxy
survives unperturbed for $\sim10$~Gyr in models $A1-B3$.
Moreover, from models $A1$, $A3$, $B1$, and $B3$, we observe that there is a negligible effect of the MW's baryonic disk on the dwarfs that are in SFDM subhalos.
 \begin{figure*}
   \centering
   \includegraphics[width=1.0\textwidth]{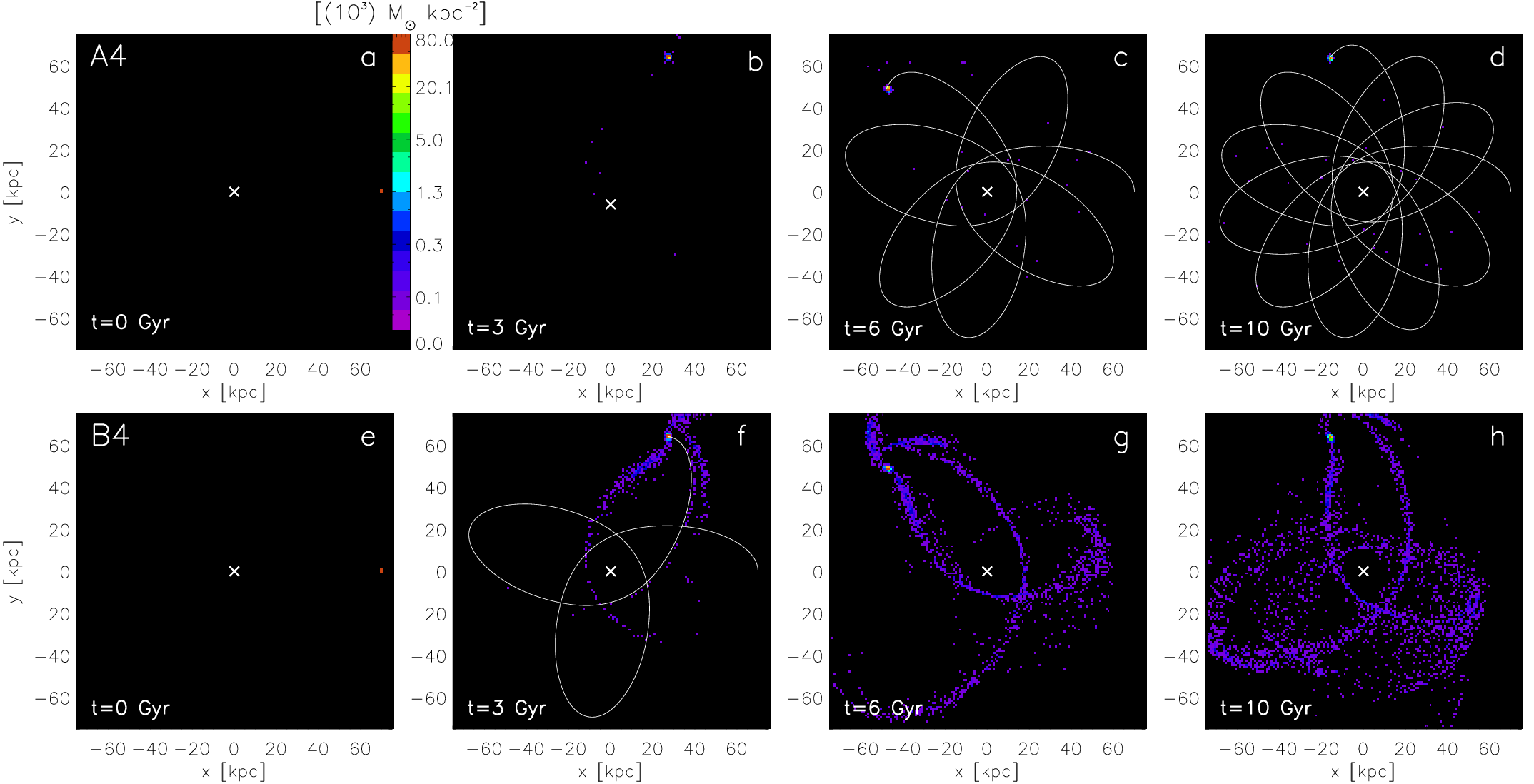}
   \caption{The surface mass density of the dwarf galaxy for $t=0$, 
  $3$, $6$, and $10$~Gyr, for models $A4$ and $B4$ centered in 
  the MW SFDM halo potential.The white cross shows the center of the MW, 
  and the white line shows the dwarf's orbit around it(colored version online).}
   \label{fig:vic4}
 \end{figure*}

The fourth pair of simulations, named as A4 and B4, resembles cases 3 but with 
$r_{p}/r_{a}=1/5$, ($A4$ and $B4$ are in Figure \ref{fig:vic2} and 
Figure~\ref{fig:vic4}). For model $A4$, the stellar component of the dwarf galaxy remains undisturbed, while the $B4$ model 
shows a major star mass loss. We run an extra couple of simulations for completeness as discussed in the next section.  
Table 1 summarizes all our simulations. 

\section{Results and Discussion}

\begin{figure*}
  \centering
  \includegraphics[width=1.0\textwidth]{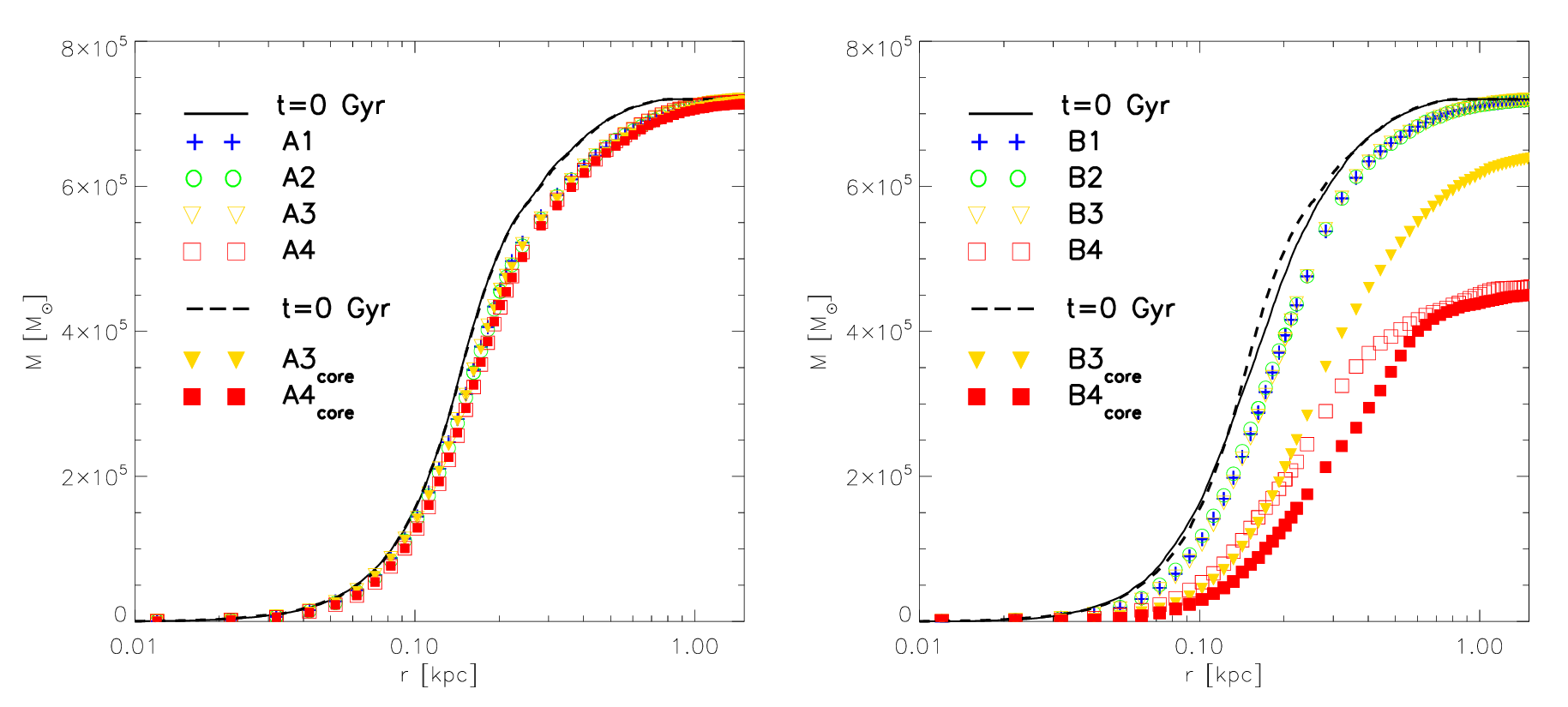}
  \includegraphics[width=1.0\textwidth]{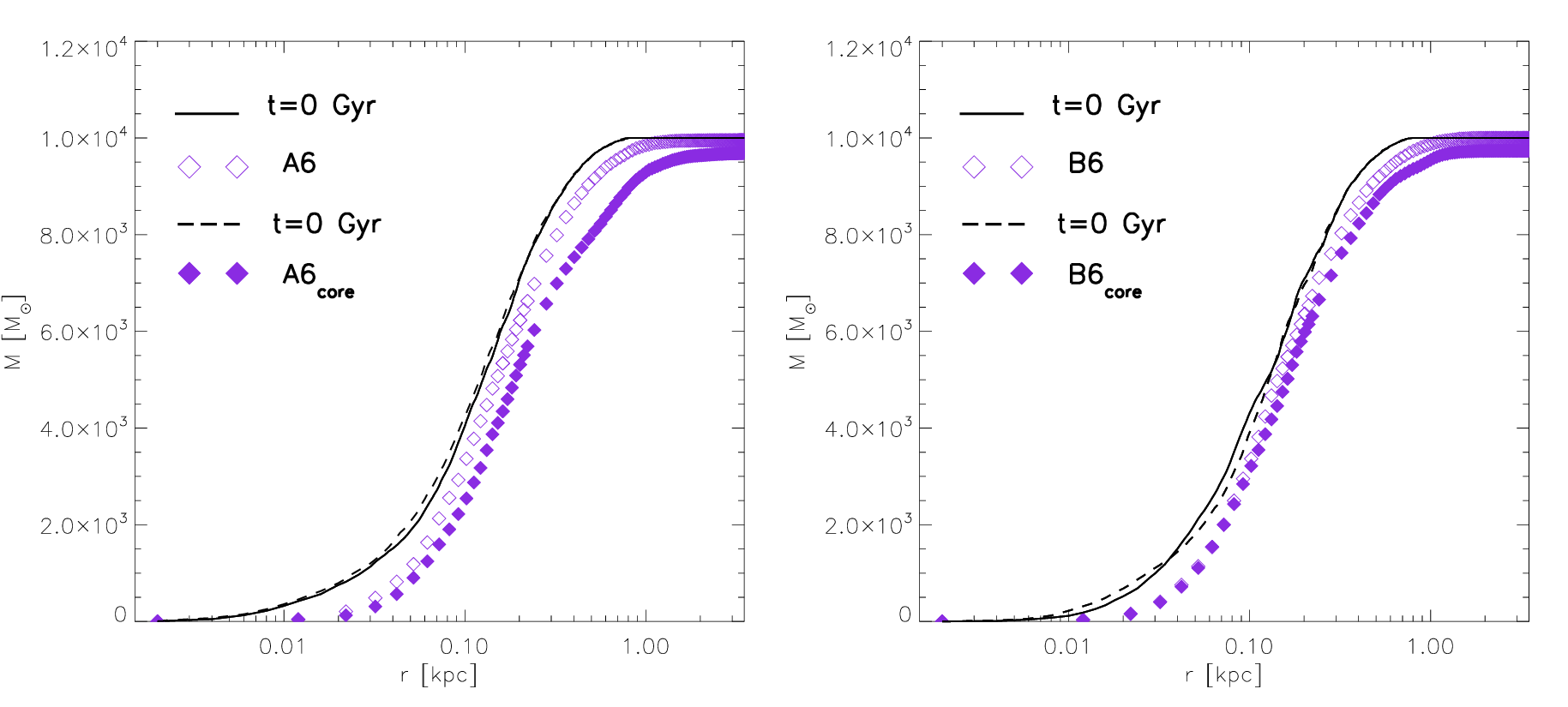}
  \caption{Upper panel: Dwarf stellar mass for models A(left) and B(right) at $t=0$ and $t=10$~Gyr. The upper left panel would represent a classical 
dwarf and the upper right would be an ultra faint-like galaxy. The different symbols in the panel represent the dark matter model used in 
that simulation according to Table 1. Bottom panels show small mass satellites with M$_{\ast}=1\times 10^4$ (models A6, A6$_{core}$, B6, and B6$_{core}$) 
at $t=0$ and $t=10$~Gyr, different symbols correspond to different simulations. In all $A$ models of the SFDM the galaxy survives at the end of the simulation independent of the stellar mass and the 
orbits we considered, even the presence of a disk in the MW scalar field halo cannot destroy the satellite. 
In B models where the subhalo is less dense, the satellite losses more mass than in A cases but will still survive inside the subhalo, 
except when the pericenter becomes comparable to the disc's scale length where we expect the scalar field subhalo to be disrupted too (color version online).}
  \label{fig:mass}
\end{figure*}

Figure~\ref{fig:mass} shows the dwarf galaxy stellar mass profile at $t=0$ and $t$ = 10~Gyr in all our simulations. 
From the upper left panel in Figure \ref{fig:mass} we see that all $A$ models lose some particles, but the loss is not substantial and 
the galaxies survive with essentially the same initial mass after $10$~Gyr.
These simulations suggest that the density is high enough to strongly bind the stars and prevent the disruption of the satellite. 
A similar behavior is seen when a cuspy-like profile is used \citep{kli09,lok12}, making tidal disruption an inefficient process 
in both core and cusp-like subhalos to reduce their stellar mass within 1kpc and therefore making it not the relevant mechanism 
that decreases the abundance of massive dwarf satellites around MW-type galaxies, even for orbits with close pericenters of 14 kpc. 

The $B$ models for the SFDM halo show a slightly larger particle loss than the $A$ models (upper right panel in Figure 6) except 
for model $B4$ which shows a more pronounced particle loss. The small central density of the SFDM dwarf subhalo, 
plays a crucial role in its survival. The final mass (at $t=10$~Gyr) for $B$ models is
smaller than the high $A$ density case in all cases. This shows that even if the orbit is far 
from the MW disk, whenever the DM subhalos have low densities the stars in the center are susceptible to spread out more 
than in denser halos as seen by comparing the two upper panels in Figure \ref{fig:mass} within 500 pc. 

One of the features that is seen from the stellar mass profiles is that the stars are not heavily stripped from the dwarf SFDM halo 
(excluding model $B4$). This is reassuring as it implies the DM density profile is also not strongly modified in that region and 
may be approximated by a fixed halo profile for orbits without small pericenters. The result is strengthened with the findings 
of \cite{pen10} for live subhalos with $r_{p}/r_{a}=1/2$ and a core DM profile; even in the presence of a live disk the DM subhalos 
remain almost the same in the central region after $10$~Gyr.
Hence, the tidal effects on the subhalo are small within $1$~kpc, which is about the relevant core size of our simulated subhalo. Therefore we consider that our approximation of a SFDM rigid halo is sufficient as long as the subhalos do not get well inside the disk of the host halo.

Simulations $A3_{core}$ and $A4_{core}$ present a similar behavior than their SFDM counterparts (see upper left panel of Fig.~\ref{fig:mass}).
In these cases the stars in the outskirts get stripped more easily, moving to larger radii and, at the same time, 
causing the inner stars to redistribute to a new configuration that follows the background DM halo potential. For cases A, 
the potential well is deep enough that only a few stars are lost; most of them remain within 1 kpc and keep the same 
initial profile.

The $B3_{core}$ model has lost more mass than its analog $B3$ 
(see the upper right panel of Fig.~\ref{fig:mass}). This is due to the slight difference
in the tail of the subhalo mass profile ($r$> 2kpc) and the fact that the potential is not as deep as in cases $A$, making it 
easier for the tidal forces to change the central stellar distribution.
From Figure 1 we note that subhalos with a core profile have a non zero density for $r$>2 kpc. For smaller $r_p:r_a$ the 
tidal stripping and the interaction with the disk becomes stronger, especially for the stars in the outermost radius which 
are more easily stripped. In fact, given that the subhalo in $B3_{core}$ is more extended than in $B3$,  
more stars are likely to get pulled toward the tail of the halo but remain inside the subhalo.
In this process the now outer stars drag some of the inner stars towards outer radii producing a more extended stellar distribution 
than in $B3$, reducing at the same time the stellar mass as shown in Fig.~\ref{fig:mass}.

The same occurs for the $B4_{core}$ model and its counterpart $B4$ (see empty and full squares in Figure~\ref{fig:mass}).
However, the considerable disruption in both $B4$ simulations indicates the need to include the disruption of the halo.
In our simulations, the satellites still remain due to the assumption of fixed subhalo, but we expect the 
dwarf halos to fully disrupt and that their stars get dispersed around the MW halo. 

We also conducted a couple of simulations ($A5$ and $B5$) where we set the dwarf galaxy embedded in the MW SFDM halo potential 
with a $r_{p}/r_{a}=1/5$ including the baryonic MW disk component and similar to $A4$ and $B4$ models, but now we place the dwarf 
galaxy in an orbit inclined $45^{\circ}$ from the $x-y$ plane. In this case we observe that the 
dwarf galaxy gets destroyed within $\sim 1.5$~Gyr. This suggests that orbits with close pericenter distances and inclination effects are factors important to the survival of low density SFDM dwarf satellites.

In order to address the dependence of our results on the stellar mass we conducted the following two pairs of simulations.
In $A6$ and $A6_{core}$, the parameters are identical to $A4$ and $A4_{core}$ respectively, but the stellar
mass of the satellite is smaller M$_{\ast}=1\times 10^4 M_{\odot}$.These parameters correspond to the closest orbit where the tidal effects should be the largest.
The other pair, $B6$ and $B6_{core}$, uses $M_{\ast}= 1\times 10^4 M_{\odot}$ and the parameters of $B3$ and $B3_{core}$ respectively.
We do not use parameters of $B4$ for the reasons mentioned in the previous paragraphs. 
These cases are shown in the bottom panels of Fig.~\ref{fig:mass}. In the bottom left of this figure we notice 
that the inner mass of the satellite reduces due to the proximity to the disk's influence but the potential well 
is again deep enough to ensure the survival of the satellite. In the bottom right panel of the same figure we see that both cases $B6$ lead 
effectively to the same result; despite their low masses the satellites can remain with most of their initial mass after 10 Gyr. 
We point out that this is consistent with the arguments given above for $B3_{core}$. In $B6_{core}$ the satellite 
has a much smaller initial stellar mass concentrated within $1$kpc, thus fewer stars are stripped during its evolution 
and are insufficient to drag most of the central stars toward the outer regions as opposed to $B3_{core}$.
Nevertheless, we still observe a small effect of this process in this pair of simulations.

Indeed, comparing cases $A$ and $B$, we notice that for the cases in which orbits are far from disk the density of the subhalo is a decisive parameter for determining the mass loss but has little influence in its survival. 
We found that the mass loss is greater if the subhalo has smaller density, but the dwarf
galaxies still survive after 10 Gyr.

When the satellites have orbits close to the center of the host or when they strongly interact with the disk, 
the subhalo central density becomes an important factor for the survival and for the number of remaining 
satellites; this is  consistent with previous works. 

It is known that in CDM simulations, the satellites with cuspy subhalos can be stripped of stars but 
still survive as DM-only subhalos, which could be detected with gravitational lensing techniques.
Here we show that if the satellites are in scalar field subhalos with central densities comparable to classical dSphs, some of their 
stars are stripped but the galaxies can survive with smaller masses and hence contribute to the number of dwarf satellites around a 
MW host. It must be noted that, in the SFDM model, 
the substructure is smaller due to the wave properties causing the 
cut-off in the power spectrum, as confirmed in \cite{sch14}. Our result can be tested with hydrodynamical SFDM cosmological simulations 
in the future. 

For the lower density dwarfs (comparable to ultra faint dwarfs), we obtained that they could survive but only if their orbits do 
not get well inside the disks of their hosts. 
On the other hand, low density halos with close pericenter orbits can be fully stripped of stars if evolved for a long time 
even with a fixed subhalo potential, but as mentioned before we expect them to be destroyed once the fixed halo hypothesis is 
relaxed. Therefore, we do not get DM halos that are tiny and dark, contrary to the CDM predictions where 
the cusp prevents total disruption. 

The formation of ultra faint dwarfs is still not clear but it is thought that they are the result of more massive dwarfs that were disrupted and left them as low density systems. We have seen that in the scenario of SFDM, depending on their distance to their host, these faint systems could also be produced from disrupted dwarfs with initial core profiles in the same way that they are when halos are assumed to have cusp profiles \citep{lok12}.

Therefore, our results point to an alternative solution to the satellite overabundance problem and the cusp-core issue by means of the quantum DM properties of the scalar field without relying strongly on the messy astrophysical processes. 
Here, small mass subhalos with core profiles ($\rho(r) \sim r^{0}$) and with orbits not crossing the host's disk are able to survive for a long time; otherwise the close encounters with the disk could completely destroy them. On the other hand, more massive dwarfs can get closer or farther from the host disk and still survive with core profiles. To determine the final fate of these galaxies and test the results from the present work, we will need simulations that involve the complexities of astrophysics, but we leave that for a future work. 


From eq. (3) we see that for larger excited states the inner region becomes more compact and the halo core sizes 
can become smaller. In general a superposition of states may be present in a SFDM halo\citep{rob13}, however, in \cite{ure10} they 
found that for a stable multi-state SFDM halo the number of bosons in excited states should decrease with increasing $j$, for instance, 
in a multi-state composed of the ground($j$=1) and first excited states ($j$=2) the number of bosons in $j$=2 should be less than or 
equal to those in $j$=1 if the halo is to be stable, hence small core sizes are expected in massive galaxies, i.e., those where the SF ground 
state provides a poor description,  depending on the required number of excited states in the halo the core size can become smaller. Given that dwarf galaxies are consistent with halos with only $j$=1, and from to the stability constraint in the SFDM halos any contribution of 
higher energy states will be subdominant in such halos and the core sizes are dictated by the dominant ground state, being of $\sim$ kpc 
for a boson mass of $\sim$10$^{-22}$ eV/$c^2$.  By fitting galaxies of different sizes it is possible to constrain the number of states in the 
SFDM halos required to agree with observations. As noted before, dwarfs seem to lie in ground state SFDM halos, while larger galaxies 
seem to require more than that, however, at this point it is unclear whether there is a mechanism that predicts the final superposition of a 
given SFDM halo; a statistical analysis fitting galaxies of different morphological types and sizes is required to derive such a relation.

Additionally, although we use a single state($j$=4) as a fit to the MW, we have not addressed its stability. 
Looking at the more general SFDM halo scenario mentioned above, it remains to be shown under what circumstances the interplay of the different 
particle states within a SFDM halo, in our case a MW-like halo, can lead to a stable halo with a non-negligible fraction of bosons 
in a single state that extends to large radii; this is something that needs to be addressed at a later time and will be vital to 
the viability of SFDM.

\section[]{CONCLUSIONS}

In this paper we explored within the context of the SFDM the influence that tidal forces have 
on an stellar distribution that is embedded in a SFDM halo with the distinctive feature of a flat central density. The satellite orbits 
around a dark matter halo with parameters that resemble those of the MW 
also modeled with a scalar field DM halo.

In general, the survival of a satellite depends on effects like its orbit, density, 
supernova explosions, its star formation history, and merger history in case that a hierarchical model is assumed among other factors.
Here we focused our analysis on the mass loss of a satellite galaxy hosted by a DM subhalo with two flat central density profiles: 
one that is predicted by the SFDM model as a consequence of the uncertainty principle, preventing the divergence of the central density,
and the second profile which is a widely used empirical model, describing core-like mass distributions, frequently encountered in the literature. We explored different central densities for the halo and concentrated on the stellar evolution under various orbits and 
different stellar masses for both halo models. 

Our results show that objects like Draco, with presumably large core sizes today, may have a simple explanation for their observed flat light distribution in the SFDM model because the stellar distribution remains bound even for tight orbits. A similar result was found in the context of CDM halos but on the condition
of choosing more specific orbits \citep{wil04,lux10}.
Also, all galaxies in our $A$ cases with core profiles survive for 10 Gyr. Thus dwarf galaxies that were accreted a long time ago will still persist as satellites with perhaps slighltly smaller mass depending on the proximity of their orbits to their hosts. A comparison with the number of satellites is out of the scope of this work because for such a comparison we need to consider the uncertainties associated to specific details of the host formation and any recent major mergers that it could have had. Simulations addressing this issue would be desirable.  

We also find a difference in the properties of the ultra faints; we showed that tidal disruption is also a mechanism for producing
ultra faint dwarfs out of more massive dwarfs. Nevertheless the disruption does not produce a substantial change in the slope of 
the inner density profile. Therefore, even with the same mechanism used in CDM to form an ultra faint dwarf there is a stark 
difference with CDM halos in their final density slopes. In fact, while cores are a direct consequence in SFDM, CDM halos rely on effective 
supernovae feedback to form cores a long time ago, otherwise they are expected to retain their divergent mass profile which may be used to differentiate these models in the future.

Regarding the too-big-to-fail issue, to make a fair comparison with CDM simulations we will need a cosmological simulation that models 
the dark matter as a scalar field in order to study the number of the most massive halos around MW-sized hosts and determine 
if this represents a challenge in the SFDM model.

We think that further exploring the SFDM looks promising due to the possible natural solutions that it proposes to long-standing issues of the CDM model. Baryonic processes should be included in SFDM simulations as they can be more directly compared with observations.

\section*{Acknowledgments}
T.M thanks Joel Primack for helpful discussions. V.L. gratefully acknowledges support from the FRONTIER grant, and HB-L.
The authors would like to thank the anonymous referee for insightful comments that helped increase the scope and improve the clarity of this paper.
This work was supported in part by DGAPA-UNAM grant IN115311 and by CONACyT Mexico grant 49865-E.
This work was partially supported by CONACyT
M\'exico under grants CB-2009-01, no. 132400, CB-2011, no. 166212,  and I0101/131/07 C-
234/07 of the Instituto Avanzado de Cosmologia (IAC) collaboration 
(http://www.iac.edu.mx/), and by the CONACyT Project no. 165584.
Victor H. Robles is supported by a CONACYT scholarship.


\end{document}